\documentstyle[aps,pra,epsfig,float,twocolumn]{revtex}
\begin{document}
\twocolumn[\hsize\textwidth\columnwidth\hsize\csname
@twocolumnfalse\endcsname

\title{Entanglement in SU(2)-invariant quantum systems:
The positive partial transpose criterion and others}

\author{John Schliemann}

\address{Department of Physics and Astronomy, University of Basel,
CH-4056 Basel, Switzerland}

\date{\today}

\maketitle

\begin{abstract}
We study entanglement in mixed bipartite quantum 
states which are invariant under simultaneous SU(2) transformations
in both subsystems. Previous results on the behavior of such states
under partial transposition are substantially extended. The spectrum of the
partial transpose of a given SU(2)-invariant density matrix $\rho$ is entirely
determined by the diagonal elements of $\rho$ in a basis of tensor-product 
states of both spins with respect to a common quantization axis. We construct a
set of operators which act as entanglement witnesses on SU(2)-invariant
states. A sufficient criterion for $\rho$ having a negative partial transpose
is derived in terms of a simple spin correlator. The same condition is
a necessary criterion for the partial transpose to have the maximum number
of negative eigenvalues. Moreover, we
derive a series of sum rules which uniquely determine the eigenvalues
of the partial transpose in terms of a system of linear equations.
Finally we compare our findings with other entanglement criteria including
the reduction criterion, the majorization criterion, and the recently
proposed local uncertainty relations.
\end{abstract}
\vskip2pc]

\section{Introduction}

As it was recognized already in the 1930's by some of the
founding fathers of modern physics, the notion of entanglement is one of the
most intriguing properties of quantum mechanics, distinguishing the quantum
world form the classical one \cite{Einstein35,Schrodinger35}. Moreover,
quantum entanglement is the key ingredient to many if not almost all 
concepts and proposal in the field of quantum information theory and
processing \cite{Nielsen00}; 
recent reviews on progress in the theoretical description and analysis of 
entanglement are listed in Refs.~\cite{Lewenstein00,Terhal02,Bruss02}.

As far as pure states of a quantum system are concerned, the
situation is, from a theory point of view, very clear: there are simple
and efficient methods to detect an quantify entanglement in a given pure
state. One of the most widely used entanglement measures for this case is 
certainly the von Neumann-entropy of partial density matrices constructed
from the full pure-state density matrix \cite{Bennett96}. However, the
problem of entanglement in mixed states is in general an open one. A mixed
state is said to be non-entangled (or separable) if it can be represented
as a convex sum of projectors onto non-entangled pure states. In the
following we shall concentrate on bipartite systems. As is was noticed by
Peres \cite{Peres96}, a necessary criterion for mixed state of a bipartite
system to be separable is that its partial transpose with respect to one
of the subsystems is positive \cite{note1}. 
Subsequently it was shown by the Horodecki
family that this condition is also sufficient if the Hilbert space
of the bipartite system has dimension $2\times 2$ or $2\times 3$
\cite{Horodecki96}. For larger dimensions, inseparable states with
positive partial transpose (PPT) exist \cite{Horodecki97}, i.e. the
PPT (or Peres-Horodecki)
criterion is in general a necessary but not a sufficient one.

More recently, mixed states being invariant under certain joint symmetry
operations of the bipartite system have been studied \cite{Vollbrecht01}.
The probably oldest example known to the literature of this kind of
objects are the Werner states. Here both parties have local Hilbert spaces
of the same dimension, and the Werner states are defined by being invariant 
under {\em all} simultaneous unitary transformations $U\otimes U$. 
Another important example from this class of states but with an in general 
much samller symmetry group are so-called SU(2)-invariant
states \cite{Schliemann03,note2}. Here we regard the two subsystems as
spins $\vec S_{1}$, $\vec S_{2}$, where $2S_{1}+1$, $2S_{2}+1$ are the 
dimensions of the corresponding Hilbert spaces.
SU(2)-invariant states are are defined to be 
invariant under all uniform rotations $U_{1}\otimes U_{2}$ of both spins
$\vec S_{1}$ and $\vec S_{2}$, where $U_{1/2}=\exp(i\vec\eta\vec S_{1/2})$
are transformations
corresponding to the same set of real parameters $\vec\eta$
in the representation of SU(2) appropriate for the spin lengths 
$S_{1}$ and $S_{2}$ ($\hbar=1$). Werner states and SU(2) invariant states are
identical for $S_{1}=S_{2}=1/2$, but for larger spin lengths the 
SU(2)-invariant states have a clearly smaller symmetry group.
By construction, SU(2)-invariant states 
commute with all components of the total spin 
$\vec J=\vec S_{1}+\vec S_{2}$. In particular, for SU(2)-invariant states
acting on bipartite Hilbert spaces with dimension $2\times N$, the 
Peres-Horodecki criterion can be shown to be necessary {\em and sufficient}
\cite{Schliemann03}, i.e. there are no entangled states of this kind 
with a positive partial transpose.

SU(2)-invariant density matrices 
arise from thermal equilibrium states of low-dimensional
spin systems with a rotationally invariant Hamiltonian 
by tracing out all degrees of freedom but those two spins. In fact, in the
recent years, entanglement in generic quantum spin models has developed to 
a major direction of research, see, e.g., 
Refs.~\cite{Osterloh02,Osborne02,Verstraete04,Wang05,note3}. Most recently, 
SU(2)-invariant states were also studied as a model for entangled
multiphoton states produced by parametric down-conversion \cite{Durkin04}.

Most recently, and during the present work was being completed, a preprint
by Breuer appeared \cite{Breuer05} where SU(2)-invariant states with
common spin length, $S_{1}=S_{2}$ are studied. The approach there is so far
restricted to small spin lengths ($S_{1}=S_{2}\leq 3/2$), but has the merit
to allow for an analysis on the sufficiency of the
Peres-Horodecki criterion.

In the present work we extend previous results on entanglement properties
of SU(2)-invariant states \cite{Schliemann03}
and compare the PPT criterion with other
entanglement criteria including the reduction criterion 
\cite{Horodecki99,Cerf99}, the majorization criterion
\cite{Nielsen01}, and the local uncertainty relations studied very recently
\cite{Hofmann03,Guhne04}. The latter criteria are very readily applied to
SU(2)-invariant states, and these considerations provide instructive
illustrations of the logical hierarchy of those entanglement criteria.

This paper is organized as follows. In section \ref{PPT} we summarize
important properties of SU(2)-invariant states under partial transposition and
derive a series of additional results which allow to extend previous findings
\cite{Schliemann03} to the case of larger spin lengths. In the following 
section we apply the above-mentioned other entanglement criteria to
SU(2)-invariant density matrices and compare the results with each other.
We close with conclusions in the last section.

\section{SU(2)-invariant states under partial transposition}
\label{PPT}

An SU(2)-invariant state $\rho$ of a bipartite system of two spins
$\vec S_{1}$, $\vec S_{2}$ has the general form \cite{Schliemann03}
\begin{equation}
\rho=\sum_{J=|S_{1}-S_{2}|}^{S_{1}+S_{2}}
\frac{A(J)}{2J+1}\sum_{J^{z}=-J}^{J}
|J,J^{z}\rangle_{0}{_{0}\langle J,J^{z}|}\,,
\end{equation} 
where the constants $A(J)$ fulfill $A(J)\geq 0$, $\sum_{J}A(J)=1$.
Here $|J,J^{z}\rangle_{0}$ denotes a state 
of total spin $J$ and $z$-component $J^{z}$. In particular, $\rho$
commutes with all components of the total spin 
$\vec J=\vec S_{1}+\vec S_{2}$. Obviously the SU(2)-invariant density matrices
from a convex set, i.e. with two given SU(2)-invariant states $\rho_{1}$,
$\rho_{2}$ any convex combination $\lambda\rho_{1}+(1-\lambda)\rho_{2}$,
$\lambda\in[0,1]$, has the same property.
Let us now consider the partial transpose
of an SU(2)-invariant state, $\rho^{T_{2}}$,  
where we take, without loss of generality,
the partial transpositions to be performed of the second subsystem describing
the spin $\vec S_{2}$. Moreover, let us assume that the partial transposition
is performed in the standard basis of joint tensor-product eigenstates of
$S^{z}_{1}$ and $S^{z}_{2}$.
As shown earlier \cite{Schliemann03}, under these conditions $\rho^{T_{2}}$ 
commutes with all components of the vector $\vec K$ defined by
$K^{x}=S^{x}_{1}-S^{x}_{2}$, $K^{y}=S^{y}_{1}+S^{y}_{2}$,
$K^{z}=S^{z}_{1}-S^{z}_{2}$, and these operators also furnish a
representation of su(2), $[K^{\alpha},K^{\beta}]=
i\varepsilon^{\alpha\beta\gamma}K^{\gamma}$ (using standard notation).
We note that the above result relies on the transformation
properties of $\rho^{T_{2}}$ \cite{Schliemann03}.  
The form of the operators $\vec K$ depends on
the basis with respect to the partial transposition is performed. 
For any choice of basis one finds a set of operators $\vec K$ commuting with
$\rho^{T_{2}}$ and fulfilling the angular momentum algebra, but the form
of the operators will in general be different from the above one obtained
in the standard basis. From the above observations it follows that 
the eigensystem of $\rho^{T_{2}}$ has the same multiplet structure
as $\rho$ \cite{Schliemann03} and can therefore be written in the general form 
\begin{equation}
\rho^{T_{2}}=\sum_{K=|S_{1}-S_{2}|}^{S_{1}+S_{2}}
\frac{B(K)}{2K+1}\sum_{K^{z}=-K}^{K}
|K,K^{z}\rangle_{0}{_{0}\langle K,K^{z}|}\,,
\label{multB}
\end{equation}
where the multiplets are labeled by the value of
$\vec K^{2}=K(K+1)$ with $|S_{1}-S_{2}|\leq K\leq S_{1}+S_{2}$ and have
degeneracy $2K+1$.
Again, the real coefficients $B(K)$ fulfill $\sum_{K}B(K)=1$
(since ${\rm tr}\rho={\rm tr}\rho^{T_{2}}$) but are not necessarily positive.
As pointed out by Peres \cite{Peres96}, negative $B(K)$ indicate entanglement
in the original state $\rho$. The coefficient of the largest multiplet,
$K=S_{1}+S_{2}$, is given by \cite{Schliemann03}
\begin{equation}
\frac{B(S_{1}+S_{2})}{2(S_{1}+S_{2})+1}
=\langle\pm S_{1},\mp S_{2}|\rho|\pm S_{1},\mp S_{2}\rangle\geq 0\,,
\end{equation} 
where $|S^{z}_{1},S^{z}_{2}\rangle$ are tensor-product eigenstates 
of $S^{z}_{1}$ and $S^{z}_{2}$. In particular, $B(S_{1}+S_{2})$ is always
non-negative and can alternatively be expressed as
\begin{equation}
\frac{B(S_{1}+S_{2})}{2(S_{1}+S_{2})+1}={\rm tr}
\left[{\tilde P}_{\vec n}(S_{1}+S_{2})\rho^{T_{2}}\right]\,,
\end{equation}
where ${\tilde P}_{\vec n}(L)$ 
is the projector onto the subspace with $\vec n\cdot\vec K=L$, and $\vec n$ is
an arbitrary unit vector.
As it follows from the above multiplet structure, each eigenvalue
of $\rho^{T_{2}}$ in the subspace with $\vec n\cdot\vec K=L+1>0$ occurs also 
exactly once in the subspace with $\vec n\cdot\vec K=L$. Thus, 
for $|S_{1}-S_{2}|\leq K<S_{1}+S_{2}$ the above relation can be generalized to
\begin{equation}
\frac{B(K)}{2K+1}={\rm tr}
\left[{\tilde P}_{\vec n}(K)\rho^{T_{2}}\right]
-\left[{\tilde P}_{\vec n}(K+1)\rho^{T_{2}}\right]\,,
\end{equation}
where the right hand side can be rewritten as
\begin{eqnarray}
 & & {\rm tr}\left[\left({\tilde P}_{\vec n}(K)
-{\tilde P}_{\vec n}(K+1)\right)\rho^{T_{2}}\right]
\nonumber\\
 & & ={\rm tr}\left[\left({\tilde P}_{\vec n}(K)
-{\tilde P}_{\vec n}(K+1)\right)^{T_{2}}\rho\right]\\
 & & ={\rm tr}\left[\left(P_{\vec n}(K)-P_{\vec n}(K+1)\right)\rho\right]\,.
\label{projdiff}
\end{eqnarray}
Here $P_{\vec n}(L)$ is the projector onto the subspace with 
$\vec n(\vec S_{1}-\vec S_{2})=L$. 
In the last equation we have used the fact that the projectors 
${\tilde P}_{\vec n}(L)$
are polynomials in the operator $\vec n\cdot\vec K$ which turns, in the
standard basis, into $\vec n(\vec S_{1}-\vec S_{2})$. However, the
expression (\ref{projdiff}) contains only the spin operators $\vec S_{1}$,
$\vec S_{2}$ and the density matrix $\rho$ itself; therefore this
expression is independent of any choice of basis,
\begin{equation}
\frac{B(K)}{2K+1}
={\rm tr}\left[\left(P_{\vec n}(K)-P_{\vec n}(K+1)\right)\rho\right]\,.
\label{eigen1}
\end{equation}
Hence any separable SU(2)-invariant density matrix fulfills
\begin{equation}
{\rm tr}\left[\left(P_{\vec n}(K)-P_{\vec n}(K+1)\right)\rho\right]\geq 0\,
\end{equation}
for $|S_{1}-S_{2}|\leq K<S_{1}+S_{2}$, while 
\begin{equation}
{\rm tr}\left[\left(P_{\vec n}(K)
-P_{\vec n}(K+1)\right)\rho\right]<0\,,
\end{equation}
indicates the presence of entanglement in the state $\rho$. Thus, when
restricting the full space of density operators to the convex submanifold
of SU(2)-invariant states, the operators $[P_{\vec n}(K)-P_{\vec n}(K+1)]$,
$|S_{1}-S_{2}|\leq K<S_{1}+S_{2}$, have the properties of entanglement
witnesses \cite{Horodecki96,Terhal00}. It is an interesting question
whether and, if so, to what extend, one can relax the restriction to 
SU(2)-invariant states with this property of the operators
$[P_{\vec n}(K)-P_{\vec n}(K+1)]$ being unaltered. 
Note also that the above operators
can, by construction,  only detect entanglement in SU(2)-invariant states
with negative partial transpose, although these operators do not fulfill the
construction recipe of decomposable entanglement witnesses \cite{Terhal02}.

Moreover, the contributions to the right hand side of Eq.~(\ref{eigen1}) 
can be expressed as
\begin{equation}
{\rm tr}\left[P_{\vec n}(K)\rho\right]
=\sum_{S^{z}_{1}-S^{z}_{2}=K}
\langle S^{z}_{1},S^{z}_{2}|\rho|S^{z}_{1},S^{z}_{2}\rangle\,.
\end{equation}
Thus, the eigenvalues of the partial transpose $\rho^{T_{2}}$ are entirely
determined by the diagonal elements of $\rho$ in a basis of tensor-product 
states of both spins with respect to a common quantization axis.
In particular, the relation
\begin{eqnarray}
\frac{B(K)}{2K+1} & = & \sum_{S^{z}_{1}-S^{z}_{2}=K}
\langle S^{z}_{1},S^{z}_{2}|\rho|S^{z}_{1},S^{z}_{2}\rangle\nonumber\\
 & & -\sum_{S^{z}_{1}-S^{z}_{2}=K+1}
\langle S^{z}_{1},S^{z}_{2}|\rho|S^{z}_{1},S^{z}_{2}\rangle
\end{eqnarray}
provides a convenient way to compute the eigenvalues of $\rho^{T_{2}}$
without explicitly solving for the zeros of a characteristic polynomial.
Below we shall encounter yet another method to determine the spectrum
of $\rho^{T_{2}}$ based on sum rules for its eigenvalues.

To gain further insight into the properties of $\rho^{T_{2}}$ consider
\begin{eqnarray}
{\rm tr}\left[\vec K^{2}\rho^{T_{2}}\right] & = &
{\rm tr}\left[\left(\vec K^{2}\right)^{T_{2}}\rho\right]
\label{trKsquared1}\\
 & = & 
{\rm tr}\left[\left(
\left(\vec S_{1}-\vec S_{2}\right)^{2}\right)\rho\right]
\label{trKsquared2}
\end{eqnarray}
for $0\leq n \leq2\min\{S_{1},S_{2}\}$. 
In the last equation we have used that the operator $S_{2}^{y}$ , 
when expressed in the standard basis, changes sign under partial 
transposition while
$S_{2}^{x}$ and $S_{2}^{z}$ remain unaltered. Alternatively, the left hand
side of Eq.~(\ref{trKsquared1}) can also be evaluated using Eq.~({\ref{multB})
leading to
\begin{equation}
\left\langle\left(\vec S_{1}-\vec S_{2}\right)^{2}
\right\rangle
=\sum_{K=|S_{1}-S_{2}|}^{S_{1}+S_{2}}K(K+1)B(K)\,,
\label{linrel1}
\end{equation}
where $\langle\cdot\rangle$ denotes an expectation value with respect to
$\rho$. 

It is instructive to investigate the condition
\begin{equation}
\left\langle\left(\vec S_{1}-\vec S_{2}\right)^{2}
\right\rangle
\geq \left(S_{1}+S_{2}\right)\left(S_{1}+S_{2}+1\right)
\label{negcrit1}
\end{equation}
which is equivalent to 
\begin{equation}
\left\langle\vec S_{1}\cdot\vec S_{2}\right\rangle<-S_{1}S_{2}
\label{negcrit2}
\end{equation}
and implies that $\rho^{T_{2}}$ has at least one negative eigenvalue,
since otherwise we had
\begin{eqnarray}
 & & \left\langle\left(\vec S_{1}-\vec S_{2}\right)^{2}
\right\rangle
=\sum_{K=|S_{1}-S_{2}|}^{S_{1}+S_{2}}K(K+1)B(K)\nonumber\\
 & &\qquad\qquad\leq\left(S_{1}+S_{2}\right)\left(S_{1}+S_{2}+1\right)
\sum_{K=|S_{1}-S_{2}|}^{S_{1}+S_{2}}B(K)\nonumber\\
 & &\qquad\qquad=\left(S_{1}+S_{2}\right)\left(S_{1}+S_{2}+1\right)\,.
\end{eqnarray}
Thus, the inequalities (\ref{negcrit1}) and (\ref{negcrit2}) are a {\em
sufficient 
condition for $\rho^{T_{2}}$ having at least one negative eigenvalue},
and, in turn, for $\rho$ being entangled. The latter statement follows also
directly from (\ref{negcrit2}), because the right hand side of this
inequality represents the minimum value the correlator 
$\langle\vec S_{1}\cdot\vec S_{2}\rangle$ can attain in a separable state.
Therefore, if (\ref{negcrit2}) is fulfilled, the underlying state must
be entangled. Note that for general spins $\vec S_{1}$, $\vec S_{2}$
the above correlator is bounded by $-(S_{1}+1)S_{2}\leq
\langle\vec S_{1}\cdot\vec S_{2}\rangle\leq S_{1}S_{2}$ (assuming
$S_{1}\geq S_{2}$). 

Moreover, the conditions (\ref{negcrit1}) and (\ref{negcrit2}) are
also a {\em necessary criterion for $\rho^{T_{2}}$ having the maximum
possible number of negative eigenvalues}. Here all $B(K)$ with
$|S_{1}-S_{2}|\leq K<S_{1}+S_{2}$ are negative, while 
$B(S_{1}+S_{2})=:\bar B>1$
because of the normalization condition $\sum_{K}B(K)=1$. 
The assertion is proved as follows,
\begin{eqnarray}
 & & \left\langle\left(\vec S_{1}-\vec S_{2}\right)^{2}
\right\rangle
\geq\left(S_{1}+S_{2}-1\right)\left(S_{1}+S_{2}\right)(1-\bar B)
\nonumber\\
 & &\qquad\qquad\qquad\qquad
+\left(S_{1}+S_{2}\right)\left(S_{1}+S_{2}+1\right)\bar B\nonumber\\
 & &\qquad\qquad\qquad=\left(S_{1}+S_{2}-1\right)\left(S_{1}+S_{2}\right)
\nonumber\\
& &\qquad\qquad\qquad\qquad+2\bar B\left(S_{1}+S_{2}\right)\nonumber\\
 & &\qquad\qquad\qquad\geq\left(S_{1}+S_{2}\right)\left(S_{1}+S_{2}+1\right)\,.
\end{eqnarray} 

The above considerations can obviously be extended to higher powers
of $\vec K^{2}$, i.e. $(\vec K^{2})^{n}$ with $n>1$. However, when performing
the partial transposition more complicated operator products occur
which give rise to additional contributions. E.g. for the next higher powers 
one finds
\begin{equation}
\left(\left(\vec K^{2}\right)^{2}\right)^{T_{2}}=
\left(\left(\vec S_{1}-\vec S_{2}\right)^{2}\right)^{2}
+4\vec S_{1}\cdot\vec S_{2}
\label{K2}
\end{equation}
and
\begin{eqnarray}
 & & \left(\left(\vec K^{2}\right)^{3}\right)^{T_{2}}=
\left(\left(\vec S_{1}-\vec S_{2}\right)^{2}\right)^{3}
-32\left(\vec S_{1}\cdot\vec S_{2}\right)^{2}\nonumber\\
 & & \qquad\qquad
+4\left(3(S_{1}(S_{1}+1)+S_{2}(S_{2}+1))-4\right)\vec S_{1}\cdot\vec S_{2}
\nonumber\\
 & & \qquad\qquad+8S_{1}(S_{1}+1)S_{2}(S_{2}+1)
\label{K3}
\end{eqnarray}
leading to the additional sum rule
\begin{eqnarray}
 & & \left\langle\left(\left(\vec S_{1}-\vec S_{2}\right)^{2}\right)^{2}
+4\vec S_{1}\cdot\vec S_{2}\right\rangle\nonumber\\
 & & =\sum_{K=|S_{1}-S_{2}|}^{S_{1}+S_{2}}\left(K(K+1)\right)^{2}B(K)
\label{linrel2}
\end{eqnarray}
and an analogous relation for $n=3$
following from Eq.~(\ref{K3}). Eqs.~(\ref{linrel1}),
(\ref{linrel2}), together with the normalization condition
$\sum_{K}B(K)=1$, form a series of sum rules being linear in the coefficients
$B(K)$. This series can obviously be extended to arbitrary high powers of the 
spin operators. The number of independent sum rules, however, is in general
given by $2\min\{S_{1},S_{2}\}+1$. Thus, for given $S_{1}$, $S_{2}$, the
relations arising from $n=0,\dots,2\min\{S_{1},S_{2}\}$ constitute a linear
system of equations which uniquely determines the spectrum of $\rho^{T_{2}}$.
Note that the coefficients in this system of equations are of the form
$(K(K+1))^{n}$, i.e. the corresponding matrix is of the Vandermonde type
with its determinant given by
\begin{equation}
\prod_{{K,L=|S_{1}-S_{2}|}\atop{K>L}}^{S_{1}+S_{2}}
(K(K+1)-L(L+1))\,,
\end{equation}
which is always positive. Such a system of linear equations for
the coefficients $B(K)$ provides an alternative way to compute
the eigenvalues of $\rho^{T_{2}}$ in terms of spin correlators.

Moreover, using the relation
\begin{equation}
\left\langle\left(\vec S_{1}+\vec S_{2}\right)^{2}
\right\rangle
=\sum_{J=|S_{1}-S_{2}|}^{S_{1}+S_{2}}J(J+1)A(J)
\end{equation}
and Eq.~(\ref{linrel1}) one derives the following sum rule
\begin{eqnarray}
 & & 2\left(S_{1}\left(S_{1}+1\right)+S_{2}\left(S_{2}+1\right)\right)
\nonumber\\
 & & \qquad=\sum_{L=|S_{1}-S_{2}|}^{S_{1}+S_{2}}
L\left(L+1\right)\left(A(L)+B(L)\right)\,,
\end{eqnarray}
where the left hand side is independent of the given state $\rho$.

Let us illustrate the above findings on some examples. The simple case when
on e of the spins, say $\vec S_{2}$, has length $1/2$ was already fully 
discussed in Ref.~\cite{Schliemann03}. Here one finds
\begin{eqnarray}
B\left(S-\frac{1}{2}\right) & = & \frac{1}{2S+1}
\left(S+2\left\langle\vec S_{1}\cdot\vec S_{2}\right\rangle\right)\,,\\
B\left(S+\frac{1}{2}\right) & = & \frac{1}{2S+1}
\left(S+1-2\left\langle\vec S_{1}\cdot\vec S_{2}\right\rangle\right)
\end{eqnarray}
where $S:=S_{1}$.
Clearly, $B(S+1/2)$ is always nonnegative (since
$\langle\vec S_{1}\cdot\vec S_{2}\rangle\leq S/2$), while $B(S-1/2)$
becomes negative if $\langle\vec S_{1}\cdot\vec S_{2}\rangle<-S/2$,
in accordance with the above results for general spin lengths. 
Moreover, as shown in Ref.~\cite{Schliemann03}, in the case $S_{2}=1/2$,
there are no entangled states with positive partial transpose, i.e. the
Peres-Horodecki criterion for separability is necessary and sufficient. 

Next let us consider $S_{2}=1$, $S_{1}=S\geq 1$. Here we can use
the relations (\ref{linrel1}),
(\ref{linrel2}) along with the normalization condition to obtain the
coefficients $B(K)$ as
\begin{eqnarray}
B(S-1) & = & \frac{1}{2S+1}\Biggl(-1
+\left\langle\vec S_{1}\cdot\vec S_{2}\right\rangle\nonumber\\
 & & \qquad\qquad
+\frac{1}{S}
\left\langle\left(\vec S_{1}\cdot\vec S_{2}\right)^{2}\right\rangle\Biggr)
\,,\\
B(S) & = & 1-\frac{1}{S(S+1)}
\left\langle\left(\vec S_{1}\cdot\vec S_{2}\right)^{2}\right\rangle\,,\\
B(S+1) & = & \frac{1}{2S+1}\Biggl(1
-\left\langle\vec S_{1}\cdot\vec S_{2}\right\rangle\nonumber\\
 & & \qquad\qquad
+\frac{1}{S+1}
\left\langle\left(\vec S_{1}\cdot\vec S_{2}\right)^{2}\right\rangle\Biggr)
\,.
\end{eqnarray}
Again the the coefficient of the largest multiplet is of course always 
nonnegative, $B(S+1)\geq 0$, while the conditions for $B(S-1)<0$ and $B(S)<0$
read
\begin{equation}
1>\left\langle\vec S_{1}\cdot\vec S_{2}\right\rangle
+\frac{1}{S}\left\langle\left(\vec S_{1}\cdot\vec S_{2}\right)^{2}\right\rangle
\,,
\label{crit1}
\end{equation}
\begin{equation}
\left\langle\left(\vec S_{1}\cdot\vec S_{2}\right)^{2}\right\rangle>S(S+1)\,,
\label{crit2}
\end{equation}
respectively. These inequalities generalize the conditions given in 
Ref.~\cite{Schliemann03} for $S=1$ to the case of general spin length $S$.
Besides, demanding that both $B(S-1)$ and $B(S)$
should be negative leads to the necessary condition
\begin{equation}
\left\langle\vec S_{1}\cdot\vec S_{2}\right\rangle<-S\,,
\label{doublecrit}
\end{equation}
and it is also easy to explicitly show from the above relations
that at least one eigenvalue
of $\rho^{T_{2}}$ has to be negative if (\ref{doublecrit}) is fulfilled,
both in accordance with our earlier general findings.

Alternatively, the coefficients $B(K)$ characterizing $\rho^{T_{2}}$ can be
expressed in terms of the quantities $A(J)$ describing $\rho$,
\begin{eqnarray}
B(S-1) & = & \frac{2S-1}{2S+1}-\frac{S-1}{S}A(S-1)\nonumber\\
 & & -\frac{2S-1}{2S+1}\frac{S+1}{S}A(S)\,,
\label{BA1}\\
B(S) & = & \frac{1}{S+1}-\frac{2S+1}{S(S+1)}A(S-1)\nonumber\\
 & & +\frac{S-1}{S}A(S)\,,
\label{BA2}\\
B(S+1) & = & \frac{1}{(2S+1)(S+1)}+\frac{S+2}{S+1}A(S-1)\nonumber\\
 & & +\frac{2}{2S+1}A(S)\,.
\label{BA3}
\end{eqnarray}
Here $A(S+1)$ has been eliminated via the normalization condition, and
the other coefficients can be expressed in terms of spin correlators 
as follows \cite{Schliemann03},
\begin{eqnarray}
A(S-1) & = & \frac{1}{S(2S+1)}\Biggl(-S-(S-1)
\left\langle\vec S_{1}\cdot\vec S_{2}\right\rangle\nonumber\\
 & & \qquad\qquad\qquad
+\left\langle\left(\vec S_{1}\cdot\vec S_{2}\right)^{2}\right\rangle\Biggr)
\,,\label{A1}\\
A(S) & = & 1 -\frac{1}{S(S+1)}
\Biggl(\left\langle\vec S_{1}\cdot\vec S_{2}\right\rangle\nonumber\\
 & & \qquad\qquad\qquad
+\left\langle\left(\vec S_{1}\cdot\vec S_{2}\right)^{2}\right\rangle\Biggr)\,.
\label{A2}
\end{eqnarray}

Moreover, most recently Breuer has investigated the case
$S_{1}=S_{2}=1$ using a different approach and concluded that for this case
the PPT criterion is necessary and sufficient, i.e. there are no entangled 
states with positive partial transpose \cite{Breuer05}. This finding
also confirms a conjecture raised recently in Ref.~\cite{Wang05}.
The question whether this is also 
true for general $S_{1}=S>1$, $S_{2}=1$ remains open. The approach of
Ref.~\cite{Breuer05} finds linear expressions for the coefficients
$B(K)$ in terms of the $A(J)$ (in the notation used here). 
Eqs.~(\ref{BA1})-(\ref{BA3}) are an example of such a linear relation
for the case of $S_{2}=1$ and general $S_{1}=S\geq 1$, while the results of
Ref.~\cite{Breuer05} are restricted to equal spin lengths
$S_{1}=S_{2}\leq 3/2$.

\section{Comparison with other entanglement criteria}

We now compare the above findings from the PPT criterion with other 
entanglement criteria. These criteria are generally weaker than the
PPT criterion, but have the merit of being very readily applied
to SU(2)-invariant states.

\subsection{The reduction criterion and the majorization criterion}
\label{reduction}

The reduction criterion \cite{Horodecki99,Cerf99}
states that if a given state $\rho$ is separable, then the operators
\begin{eqnarray}
 & \rho_{1}\otimes {\bf 1}-\rho & \nonumber\\
 & {\bf 1}\otimes\rho_{2}-\rho & \nonumber
\end{eqnarray}
are also positive, i.e. do not contain any negative eigenvalue. Here
$\rho_{1/2}={\rm tr}_{2/1}(\rho)$ denotes the reduced density matrices of
each subsystem. This criterion is in general weaker than the PPT criterion.
If one of the subsystems has a Hilbert space of dimension two, however,
both criteria are equivalent \cite{Cerf99}. Thus, in particular,
the reduction criterion is necessary and sufficient for the case of the
dimensions $2\times 2$ or $2\times 3$ .

Applying the reduction criterion to SU(2)-invariant states is technically
very easy since, due to the rotational invariance of these objects,
we have
\begin{equation}
\rho_{1/2}=\frac{1}{2S_{1/2}+1}{\bf 1}\,.
\end{equation}
Thus, the criterion is violated if
\begin{equation}
\frac{A(J)}{2J+1}>\frac{1}{2S_{1/2}+1}
\end{equation}
for some $J$. If this inequality is fulfilled, the underlying SU(2)-invariant 
state $\rho$ is inseparable. Because $A(J)\leq 1$ this is only possible for
$J<S_{1/2}$, which strongly restricts the power of this
entanglement criterion as applied to SU(2)-invariant states \cite{Breuer05}.

Let us now compare the reduction criterion with the results obtained from
the PPT criterion. For the case $S_{1}=S$, $S_{2}=1/2$ the reduction
criterion is violated if
\begin{equation}
A(S-1/2)>\frac{2S}{2S+1}\,,
\end{equation}
or, using $A(S-1/2)=(S-2\langle\vec S_{1}\cdot\vec S_{2}\rangle)/(2S+1)$
(cf. Ref.~\cite{Schliemann03}),
\begin{equation}
\left\langle\vec S_{1}\cdot\vec S_{2}\right\rangle<-\frac{S}{2}\,.
\end{equation}
This condition is of course the same as found from the PPT criterion
since both criteria are equivalent for this case.

For the case $S_{1}=S\geq 1$, $S_{2}=1$ violation of the reduction
criterion leads to the condition
\begin{equation}
A(S-1)>\frac{2S-1}{2S+1}\,,
\end{equation}
or, using Eq.~({\ref{A1}), 
\begin{equation}
-\frac{(S-1)}{S}\left\langle\vec S_{1}\cdot\vec S_{2}\right\rangle
+\frac{1}{S}\left\langle\left(\vec S_{1}\cdot\vec S_{2}\right)^{2}\right\rangle
>2S\,.
\label{red}
\end{equation}
For $S=1$ this inequality is the same as the criterion (\ref{crit2}). The other
inequality (\ref{crit1}), however, is not reproduced by the reduction
criterion, and for $S>1$ the above inequality (\ref{red}) is a
weaker criterion for entanglement than (\ref{crit2}). 
In fact, demanding that $\rho^{T_{2}}$ is positive, i.e. $B(S-1)\geq 0$ and
$B(S)\geq 0$, 
one derives from Eqs.~(\ref{BA1}),(\ref{BA2}) the necessary condition
\begin{equation}
A(S-1)\geq\frac{S^{2}-(S-1)}{S^{2}}\frac{2S-1}{2S+1}\,.
\end{equation}
Thus, whenever the PPT criterion is unable to detect entanglement
in a given state $\rho$, the reduction criterion will also fail. 
As mentioned above, this is a general property \cite{Horodecki99,Cerf99}.

Another entanglement criterion related to the reduction criterion is the
majorization criterion \cite{Nielsen01,Bruss02}. 
It states that any separable state $\rho$ fulfills the inequalities
\begin{eqnarray}
\lambda^{\downarrow}_{\rho} 
& \prec & \lambda^{\downarrow}_{\rho_{1}}\nonumber\\
\lambda^{\downarrow}_{\rho} 
& \prec & \lambda^{\downarrow}_{\rho_{2}}\nonumber
\end{eqnarray}
where the vector $\lambda^{\downarrow}_{\rho}$ consists of the eigenvalues
of $\rho$ in decreasing order The notation $x\prec y$ means
$\sum_{j=1}^{k}x_{j}\leq\sum_{j=1}^{k}y_{j}$ for $k\in\{1,\dots,d\}$,
where the equality holds for $k=d$. Here $d$ is the dimension of the
total Hilbert space of the bipartite system, and the vectors
$\lambda^{\downarrow}_{\rho_{1/2}}$ are extended by zeros in order to make
their dimension equal to that of $\lambda^{\downarrow}_{\rho}$. Obviously the
majorization criterion is fulfilled if
\begin{equation}
\frac{A(J)}{2J+1}<\frac{1}{2S_{1/2}+1}\,.
\end{equation}
Thus, when applied to SU(2)-invariant states, the majorization criterion is 
always weaker than the reduction criterion.

\subsection{Local uncertainty relations}
\label{LUR}

Entanglement criteria based on so-called local uncertainty relations where
introduced recently by Hofmann and Takeuchi \cite{Hofmann03}, and by G\"uhne
\cite{Guhne04}. This concept is based on the following observation. Let
$\rho=\sum_{k}p_{k}\rho_{k}$, $p_{k}\geq 0$, $\sum_{k}p_{k}=1$ be a convex
combination of some states $\rho_{k}$ and let $M_{i}$ some set of 
operators. Then the following inequality holds \cite{Hofmann03,Guhne04}
\begin{equation}
\sum_{i}\delta^{2}(M_{i})_{\rho}\geq\sum_{k}p_{k}
\sum_{i}\delta^{2}(M_{i})_{\rho_{k}}
\end{equation}
where
\begin{equation}
\delta^{2}(M)_{\rho}=\left\langle M^{2}\right\rangle_{\rho}-
\left\langle M\right\rangle_{\rho}^{2}
\end{equation}
and $\langle\cdot\rangle_{\rho}$ denotes an expectation value with respect to
$\rho$. Consider now a bipartite system with operators $M_{i}^{(1)}$,
$M_{i}^{(2)}$ acting on one of the subsystems. Then for any separable state
$\rho_{k}$ one has 
\begin{equation}
\delta^{2}(M_{i}^{(1)}+M_{i}^{(2)})_{\rho_{k}}
=\delta^{2}(M_{i}^{(1)})_{\rho_{k}}+\delta^{2}(M_{i}^{(2)})_{\rho_{k}}\,.
\end{equation}
Let now $U_{1/2}$ be the absolute minimum of 
$\sum_{i}\delta^{2}(M_{i}^{(1/2)})$ with respect to all possible states
of each subsystem. 
Then any separable state $\rho$ has to
fulfill the inequality
\begin{equation}
\sum_{i}\delta^{2}(M_{i}^{(1)}+M_{i}^{(2)})_{\rho}\geq U_{1}+U_{2}\,.
\end{equation}
Violation of this inequality is indicative of entanglement in the
underlying state $\rho$. Note that this observation provides a whole variety
of entanglement criteria since the operators $M_{i}^{(1/2)}$ are undetermined
so far. In circumstances of SU(2)-invariant states, however, it is
natural to choose $M_{i}^{(1/2)}=S_{1/2}^{i}$, $i\in\{x,y,z\}$ with
$U_{1/2}=S_{1/2}$ \cite{Hofmann03}. Then a given SU(2)-invariant state
$\rho$ is entangled if
\begin{eqnarray}
\left\langle\vec S_{1}\cdot\vec S_{2}\right\rangle & < & 
-\frac{1}{2}\left(S_{1}^{2}+S_{2}^{2}\right)\\
 & = & -S_{1}S_{2}-\frac{1}{2}\left(S_{1}-S_{2}\right)^{2}\,.
\end{eqnarray}
The second version of this inequality suggests that this entanglement criterion
is strongest if both spins are of the same length, $S_{1}=S_{2}$. In this
case, the criterion again states that the
correlator $\langle\vec S_{1}\cdot\vec S_{2}\rangle$ has to be smaller
than its minimum value in any separable state, from which it follows
that the underlying state has a negative partial transpose. 

The local
uncertainty relation of the above form is based on a very natural choice
of operators, but provides in general only a quite weak entanglement
criterion. For instance, the above spin correlator is bounded from below
by $-(S_{1}+1)S_{2}\leq\langle\vec S_{1}\cdot\vec S_{2}\rangle$
(assuming $S_{1}\geq S_{2}$). Thus, the above inequality cannot be fulfilled
if $S_{1}$ sufficiently exceeds $S_{2}$. We leave it open whether another
choice of operators could lead to stronger criteria for
inseparability.

\section{Conclusions}
\label{conclusions}

We have investigated  entanglement in SU(2)-invariant bipartite quantum 
states and have substantially extended previous results on the behavior 
of such states under partial transposition. The spectrum of the
partial transpose of a given SU(2)-invariant density matrix $\rho$ is entirely
determined by the diagonal elements of $\rho$ in a basis of tensor-product 
states of both spins with respect to a common quantization axis. 
We have construct a
set of operators which act as entanglement witnesses on SU(2)-invariant
states, and we have derived 
sufficient criterion for $\rho^{T_{2}}$ having at least one negative
eigenvalue in terms of a simple spin correlator. The same condition is
a necessary criterion for the partial transpose to have the maximum number
of negative eigenvalues. Moreover, we have presented
a series of sum rules which uniquely determine the eigenvalues
of the partial transpose in terms of a system of linear equations.
Finally we have compared our findings with 
other entanglement criteria including
the reduction criterion, the majorization criterion, and the recently
proposed local uncertainty relations.

The key challenge for future investigations of SU(2)-invariant states
(or states being invariant under other transformation groups) is
certainly to determine to what extend the PPT criterion is neccessary
{\em and sufficient}. A possible route toward this goal could be
given by the methods developed in the recent preprint \cite{Breuer05}.
This approach, however, is so far limited to the case of equal  
spin lengths $S_{1}=S_{2}\leq 3/2$.


\begin{references}
\bibitem{note1}
We adopt the somewhat sloppy but common jargon that an operator is said to be
positive if it does not have any negative eigenvalue.
\end{references}
\end{document}